\documentclass[12pt,preprint]{aastex}
\usepackage{graphicx} 

\newcommand{\arc}{\ensuremath{^{\prime\prime}}}

\shorttitle{Steady Heating Model for Active Region Core}
\shortauthors{Winebarger, Warren \& Falconer}

\begin{document}


\title{Modeling X-ray Loops and EUV "Moss" in an Active Region Core}

\author{Amy R. Winebarger} 
\affil{Alabama A\&M University, 4900 Meridian Street, P.O. Box 1268, Normal, AL 35762; winebarger@physics.aamu.edu}
\author{Harry P. Warren} 
\affil{Naval Research Laboratory, Washington, DC 20375; hwarren@nrl.navy.mil}
\author{David A. Falconer}
\affil{Marshall Space Flight Center, SD50, Space Science Department, 
	Huntsville Al 35812; david.falconer@msfc.nasa.gov}


\begin{abstract}

The Soft X-ray intensity of loops in active region cores and corresponding footpoint, or moss, intensity observed
in the EUV remain steady for several hours of observation.  The steadiness of the emission
has prompted many to suggest that the heating in these loops must also be steady, though
no direct comparison between the observed X-ray and EUV intensities and the steady heating solutions of the hydrodynamic 
equations has yet been made.  In this paper, we perform these simulations and   
simultaneously model the X-Ray and EUV moss intensities in one active region core with steady uniform
heating.  To perform this task, we introduce a new technique to constrain the model parameters 
using the measured EUV footpoint intensity to infer a heating rate.  
Using an ensemble of loop structures derived from magnetic field extrapolation of photospheric field, 
we associate each field line with a EUV moss intensity, then 
determine the steady uniform heating rate on that field line that reproduces the observed EUV intensity within 5\% for a
specific cross sectional area, or filling factor.
We then calculate the total X-ray filter intensities from all loops in the ensemble and compare this to the
observed X-ray intensities.  We complete this task iteratively to determine the filling factor that returns the 
best match to the observed X-ray intensities.
We find that a filling factor of 8\%  and loops that expand with height provides the best
agreement with the intensity in two X-ray filters, though the simulated 
SXT Al12 intensity is  147\% the observed intensity and the
SXT AlMg intensity is 80\% the observed intensity.
From this solution, we determine the required heating rate
scales as ${\bar{B}}^{0.29} L^{-0.95}$.  Finally we discuss the future potential of this type of modeling, such as
the ability to use density measurements to fully constrain filling factor, and its
shortcomings, such as
the requirement to use potential field extrapolations to approximate the coronal field. 
\end{abstract}



\section{Introduction}

Determining the timescale of the heating in the solar corona is an important clue to the coronal 
heating mechanism.  One strong argument for time independent (steady) heating is the steady emission of observed X-ray 
loops in the cores of active regions (e.g., \citealt{yoshida1996}).   The footpoints of these X-ray loops
form the reticulated pattern in EUV images typically called ``moss'' (e.g., \citealt{berger1999,fletcher1999}.)
Though the intensity of the moss can vary significantly on short time and spatial scales, it
is believed that most of the variation comes from spicular material that is embedded
within the moss that can obscure and absorb the moss emission (\citealt{depontieu1999}).  The moss
intensity  averaged over  small regions is constant for many hours of observations (\citealt{antiochos2003}),
further corroborating the hypothesis that the heating in these core regions is steady.

Due to the abundance of loops in the core region, individual loop properties cannot be extracted
making it difficult to compare these regions directly with steady heating solutions of the hydrodynamic
equations.  Instead, average values for the moss intensities have been compared to the 
expected intensity for the footpoints of ``typical'' hot loops.  For instance, \cite{martens2000} found that typical moss
intensities matched expected intensities for the footpoints of multi-million degree loops when a 10\% filling
factor was included.  Such a filling factor is widely accepted for hot X-ray loops (e.g., \citealt{porter1995}).

Because individual loop properties cannot be extracted, an active region core must be modeled as an
ensemble of loops.   
There have been several recent attempts to forward model the entire solar corona (\citealt{schrijver2004}),
active regions (\citealt{lundquist2006, warren2006, warren2006b}), or bright points
(\citealt{brooks2007}) using ensembles of loops that each
satisfy the one-dimensional hydrodynamic equations.
In these attempts, the volumetric heating rate, $E$, was assumed to be a function of average 
magnetic field strength, $\bar{B}$, and loop length, $L$, i.e., $ E = E_0 \bar{B}^\alpha/L^\beta$, where different values of $E_0$, 
$\alpha$ and $\beta$ were considered.   
The formalism for the heating equations was suggested by
\cite{mandrini2000} who determined that different heating mechanisms would release energy 
as a function of the loop length and average magnetic field strength along a loop.
In the previous ensemble studies, it was found that the X-ray intensity could be well matched
by the simulations $\alpha = \beta = 1$ .  The EUV intensity, however, was poorly matched in
the case of the whole corona (\citealt{schrijver2004}) and active regions (\citealt{warren2006}).
(Note that \cite{schrijver2004} considered the heat flux through the base
of the loop, $F$, as a function of the magnetic field at the base, $B_0$, instead of the 
volumetric heating rate i.e., $F \sim B_0^\alpha/L^\beta$.  Because $E = F/L$ and
$\bar{B} \sim B_0/L$, Schrijver's reults are consistent with the other studies 
for $\alpha = 1$.)  

These previous analyses attempted to match the distribution of the total intensity of the entire
Sun or in an active region.  They included in their comparisons the long EUV loops
that are not in hydrostatic equilibrium (e.g., \citealt{lenz1999}) and are believed to be evolving
(see \citealt{winebarger2003, warren2003}).  When \cite{warren2006} compared just the simulated moss intensities to
the observed moss intensities, they found approximate agreement for heating rate scaling as $\bar{B}/L$ and 
with loops expanding with height.

In this analysis, we compare, for the first time, both the EUV moss and X-ray intensities 
to the solutions of one-dimensional hydrodynamic equations for steady uniform heating.
We use a new approach to simultaneously match the EUV moss intensity
of an active region core and the total X-ray intensity in two filters.  We make no a priori assumption
about the relationship between the heating rate and the magnetic field strength and loop length.  
Instead we use the moss intensity at the loop footpoint to find the best heating magnitude for each individual 
loop.  This approach is based on the determination that the
moss intensity observed in a narrow-band EUV filter is linearly proportional to the loop
pressure multiplied by a filling factor if the heating in the loop is steady and uniform (\citealt{martens2000}).  
The pressure combined with the loop length (determined from magnetic field extrapolation)
defines explicitly the heating rate and apex temperature for the loop 
(\citealt{rosner1978,serio1981}).

In this paper, we analyze EUV and X-ray emission of Active Region 9107 over 12 hours
on 31 May 2000.  This active region has a core region that is bright in X-ray images with
large patches of moss that are unobscured by overlying loops in EUV images. We find that the
intensity in the region varies little over the 12 hours of observation.
We determine the loop geometry using potential field extrapolations of photospheric 
field measurements. We select the loop footpoints using the moss intensity as a proxy. 
We populate the loops by assuming a filling factor, then using the observed EUV intensities at 
one footpoint to constrain the heating rate.  
We then compare the total intensity in the X-ray filters to the observed intensities.
We complete this process iteratively for different filling factors until the simulated X-ray intensities
well match the observed X-ray intensities.  We have completed this process for two different assumptions of the
cross sectional area of the loop. First
we force the cross sectional area of the loop to remain constant as supported by loop observations (\citealt{klimchuk2000});
second, we allow the cross sectional area of the loop to expand with height as $B(s)^{-1}$.
We find no satisfactory solution can be obtained for loops with constant cross sectional area.   For loops
that expand with height, a filling factor of 8\% was found to produce acceptable agreement between
the observed and simulated X-ray intensities in both filters.  From this set of solutions, we
determine that the heating scales like $\bar{B}^{0.29}/L^{0.95}$.  Finally we discuss future implications
of this method of simultaneous modeling.


\begin{figure}[t!]
\centerline{
\includegraphics{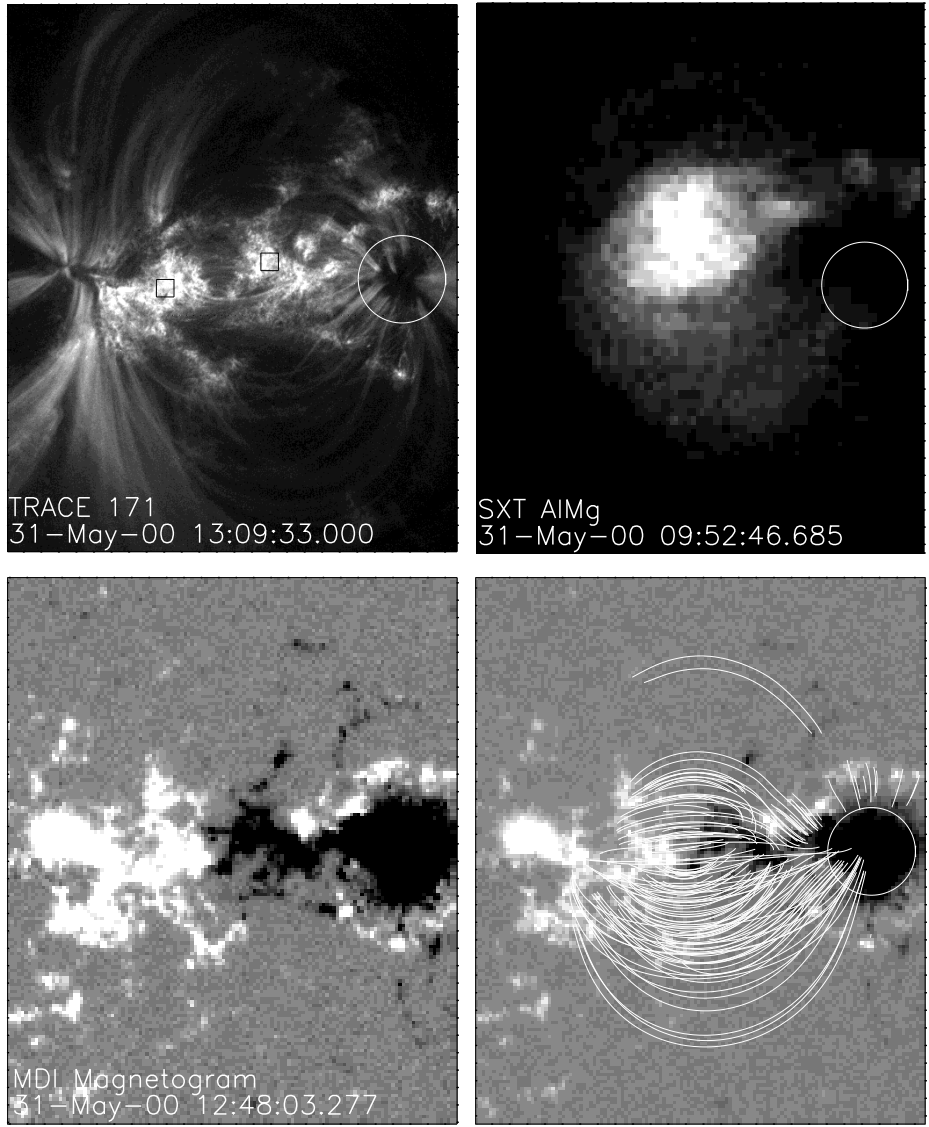}}
\caption{{\it Top row:} TRACE 171\,\AA, and SXT AlMg images of the active region.  {\it Bottom row:}
Co-aligned MDI magnetogram and field lines determined from potential field extrapolations.
\label{fig:multipanel}}
\end{figure}

\begin{figure}[t!]
\centerline{
\includegraphics{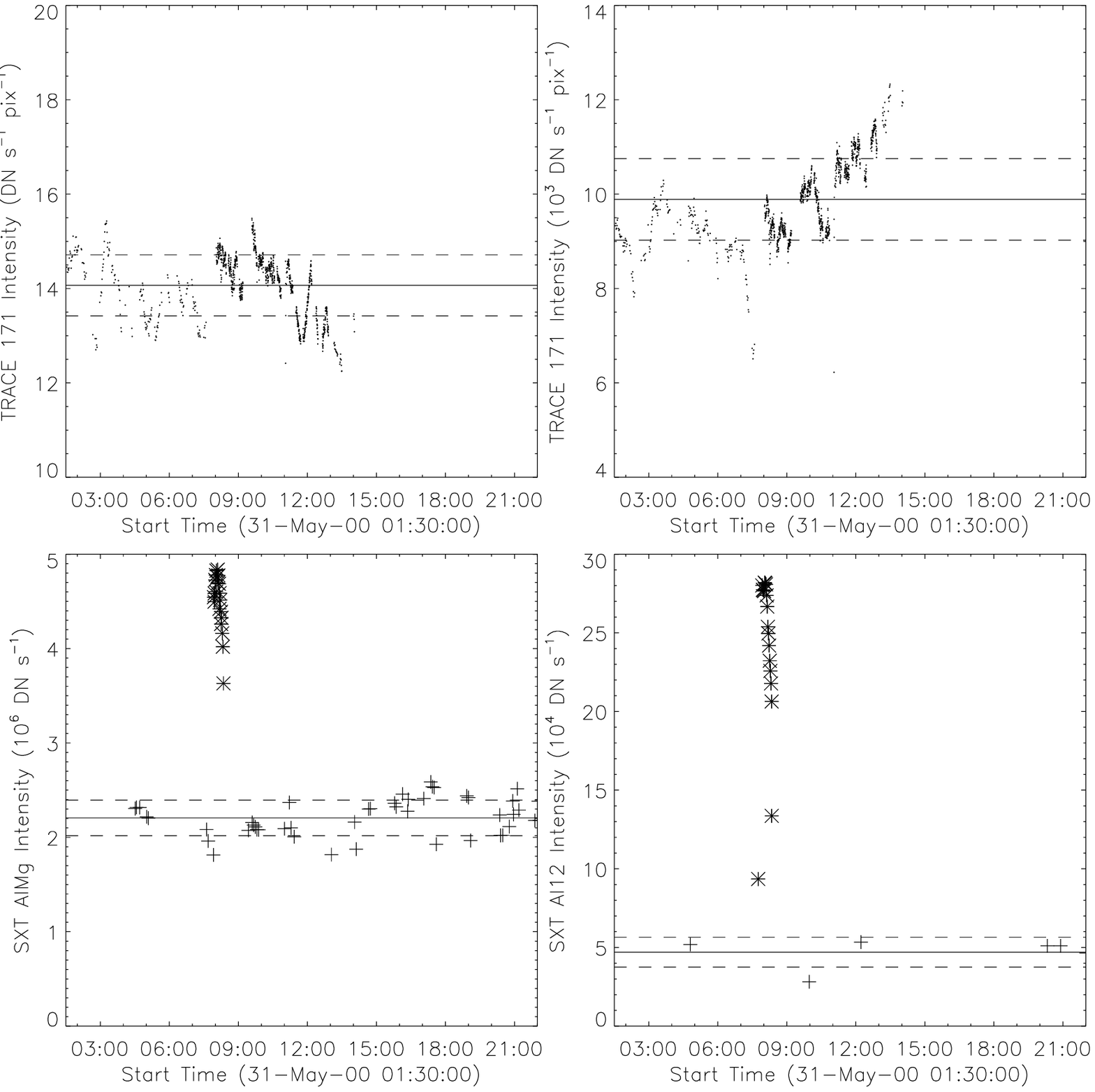}}
\caption{Top: The evolution of the TRACE 171 intensity summed over the boxes shown in Figure 1 as a function of time.  Bottom: 
The evolution of SXT AlMg and Al12 intensity summed over the entire active region. In all images, the solid line shows the average
intensity, the dashed lines show the average $\pm$ one standard deviation.  The SXT data points represented with an asterisk were
not considered in the average.
\label{fig:time}}
\end{figure}

\section{Data Analysis}

The goals of this study were to simultaneously model an Active Region core
observed in both the Soft X-ray Telescope (SXT) (\citealt{tsuneta1991}) flown
on the {\it Yohkoh} satellite and
the {\it Transition Region and Coronal Explorer (TRACE)} 
(\citealt{handy1999, schrijver1999}).  To find an adequate active region, the
available data bases were browsed for long term, simultaneous observations
of a relatively simple, non-flaring active region with an unobscured moss region 
observable in the EUV.  We selected Active Region 9017 
observed for several hours with both instruments on 2000 May 31.
We use the {\it TRACE} 171\,\AA\ filter which is sensitive to the Fe~IX/X lines formed at $\sim 1$\,MK
and the  SXT Al, Mg, Mn, C ``sandwich'' filter (AlMg) and the thick Al (Al12) filter  in this study.  
The active region is shown in Figure~\ref{fig:multipanel}.  

\subsection{Temporal Evolution}

The evolution of the TRACE
intensities averaged over the two regions of moss shown on the
TRACE 171 image in Figure~\ref{fig:multipanel} are shown on the top two panels of
Figure~\ref{fig:time}.  The average intensity in the left hand box is 14.0$\pm$ 0.6 DN~s$^{-1}$~pix$^{-1}$ and
in the right hand box is 9.9 $\pm$ 0.9 DN~s$^{-1}$~pix$^{-1}$.  The errors given are one standard 
deviations in the mean values and are less than 10\% of the average values in both cases meaning the
intensity in these moss regions remain relatively constant over the 12.5 hours of observation.

The evolutions of the total intensity summed over the active region observed with the SXT AlMg and 
Al12 filters are shown in the bottom two panels of Figure~\ref{fig:time}.
The active region brightens significantly in both filters between 7:30 and 9:00 UT due to the evolution of two 
loops which have previously been studied (\citealt{winebarger2005}).  Excluding this time frame from consideration, 
the average of the total intensity was calculated and is shown as a solid line.  In the AlMg filter, the
average intensity is 2.2 $\pm  0.2 \times 10^{6}$ DN~s$^{-1}$ and 4.7 $\pm 1.0 \times 10^{4}$ DN~s$^{-1}$ in the
AlMg and Al12 filters, respectively.  The standard deviation in the mean value of the AlMg intensity
is less than 10\%, while the standard deviation in the Al12 intensity is 20\%. 

Excluding the time frame of the evolving loops, this active region core remains relatively steady for 12.5+ hours
of observation.  For the remainder of the paper, we use
the {\it TRACE} 171\AA\ image taken at 13:09:33 UT to be representative of the moss intensity
and the average SXT AlMg and SXT Al12 intensities for comparison.


\subsection{Determining Loop Geometry}

Because it is difficult to extract information about individual loops from the observations, we use
potential field extrapolations of the co-aligned photospheric field measurements  to approximate the loop geometry.
We use photospheric field measurements from the Michelson Doppler Imager (MDI, \citealt{scherrer1995}) to estimate
the coronal field.   
The MDI magnetogram used in this study was taken at 12:48:00 UT and is shown in the second row of Figure~\ref{fig:multipanel}.
We also examined a vector magnetograph from the Marshall 
Space Flight Center Vector Magnetograph (MSFCVM, \citealt{hagyard1982}) taken
at 16:09 UT.  From the vector magnetograph, we determined that the active region was well approximated as potential.

To select the footpoints of the loops, we use the moss region of the TRACE 171 image as a proxy.   We first
select all pixels within the moss regions.
These regions
were identified visually.  We use the center of the TRACE pixels within these regions as the starting points to
trace the field lines.  If we use all the TRACE pixels in these regions as starting points, we would have over
30,000 field lines.  To reduce the number to a more manageable amount, 
we bin the TRACE images to a resolution of 1.5\arc; the result being one field line for each 9 TRACE high
resolution pixels or a total of about 3,000 field lines.  
We trace the field lines from a height of 2.5\,Mm above the solar surface which is the approximately the measured
height of the moss above the limb (\citealt{martens2000}).  Starting at this height also circumvents the problem of having 
locally closing field lines that never reach coronal heights (\citealt{warren2006}).
After tracing all the field lines, we compare the terminating footpoints and remove any field lines that are duplicates. 
A few representative field lines are shown Figure~\ref{fig:multipanel}.  
Note that several of the calculated field lines terminate in the negative polarity sunspot where there is 
no moss or SXT emission.  This region is highlighted
with a circle in Figure~\ref{fig:multipanel}.

\section{Modeling the Loops}

Before describing the details of the modeling portion of this research, it is useful to 
consider the implications of the applicable scaling laws.
\cite{martens2000} determined that the intensity in the {\it TRACE} 171\AA\  filter was linearly proportional
to the base pressure in the loop times a filling factor, i.e., 
\begin{equation}
p_0 f = 0.050 I_{171} 
\end{equation}
where $p_0$ is the base pressure in dyne cm$^{-2}$, $f$ is the volumetric filling factor, and
$I_{171}$ is the TRACE 171 intensity in DN s$^{-1}$ pixel$^{-1}$.  If we consider only a single TRACE pixel,
the filling factor in the above equation would then be related to the cross sectional
area of the loop.  A filling factor of unity would imply that the cross sectional area of the loop was at least
equal to the TRACE pixel area.  A filling factor of less than 1 would imply that the cross sectional area of the
loop was a fraction of the TRACE pixel size.

The above relationship can then be combined with the RTVS scaling laws for uniform heating
(\citealt{rosner1978,serio1981}) , i.e.,
\begin{eqnarray}
T_{max} \sim 1.4 \times 10^3 (p_0 L)^\frac{1}{3} \exp[0.04L/s_p] \\
E_H \sim 9.8 \times 10^4 p_0^\frac{7}{6} L^{-\frac{5}{6}} \exp[-0.5L/s_p]
\end{eqnarray}
where $T_{max}$ is the maximum temperature in the loop in Kelvin,  $L$ is the half length
of the loop in cm, $E_H$ is the volumetric heating rate of the loop in ergs cm$^{-3}$ s$^{-1}$,
and $s_p$ is the gravitational scale height (generally 47 Mm/MK).
These scaling laws were derived using a simplified radiative loss function and assuming
the loop semi-circular and perpendicular to the solar surface.  

For a single loop, if the moss intensity and loop length were known, the only free parameter in the
above set of equations is the filling factor.  Hence, if other constraining measurements, such as X-ray intensities
were also available, it would be possible to calculate the solutions for various filling factors and determine
which filling factor satisfactorily reproduced the X-ray intensities.  If no filling factor could be
found to return the X-ray intensities, we would conclude that the loop could not be modeled with steady, uniform
heating.

This is exactly the test we wish to perform in this study, but 
instead of a single loop, we consider an ensemble of such loops.  We follow the same process outlined above 
of using the moss intensity to constrain the heating rate on each individual loop for a given filling factor and we further
assume that all loops in the bundle have the same filling factor.  Instead of using the scaling 
laws given above, however, we solve the steady-state hydrodynamic equations that include the geometry for each loop.  
For each filling factor, we solve the equations for all the loops in the bundle, then calculate the total X-ray intensity of
all the loops  and compare this value to the observed X-ray intensity in both SXT filters.    
 We  complete this process twice, first assuming all loops in the bundle have a constant cross sectional area
and then assuming that each loop expands proportional to the inverse of its field strength.

For each field line and filling factor, we use the relationships suggested by \cite{martens2000} and 
\cite{rosner1978} to get a first approximation for the heating rate in the loop.  Because these previous works depended
on some simplifications and approximations to solve the one dimensional hydrodynamic equations,
the heating rate found is only a rough guess to the heating rate 
required to match the observed EUV intensities.
Using this guess, we then compute the solution to the hydrodynamic equations using a numerical code 
created by Aad van Ballegooijen (\citealt{hussain2002, schrijver2005}).   This code allows for the
loop geometry and area expansion to be included in the solution.   (Note that we use the radiative loss function calculated
by \cite{brooks2006}; all instrument response functions were calculated using the same atomic data and are fully
consistent with the radiative loss function and one another.) After computing the solution, we fold the density and temperature
through the {\it TRACE} 171 filter response function, compute the simulated moss intensity, and compare
the simulated moss intensity to the observed moss intensity associated with that
fieldline. If the simulated moss intensity is too low, we increase the heating rate; if the simulated moss intensity
is too high, we decrease the heating.  We complete this procedure iteratively until the 
simulated moss intensity is within 5\% of the observed moss intensity.  The result is the uniform heating
rate and resulting density and temperature on the field line that reproduces the moss intensity associated with the footpoint.
After the best hydrodynamic solution for each field line has been computed, we then 
calculate the resulting SXT AlMg and Al12  by convolving the density and temperature with the
SXT filter response functions.  
We sum the SXT filter intensities from all the loops in the ensemble then compare it with the observed AlMg and Al12 
intensities.

\section{Results}

\begin{figure*}[t!]
\centerline{
\resizebox{18cm}{!}{\includegraphics{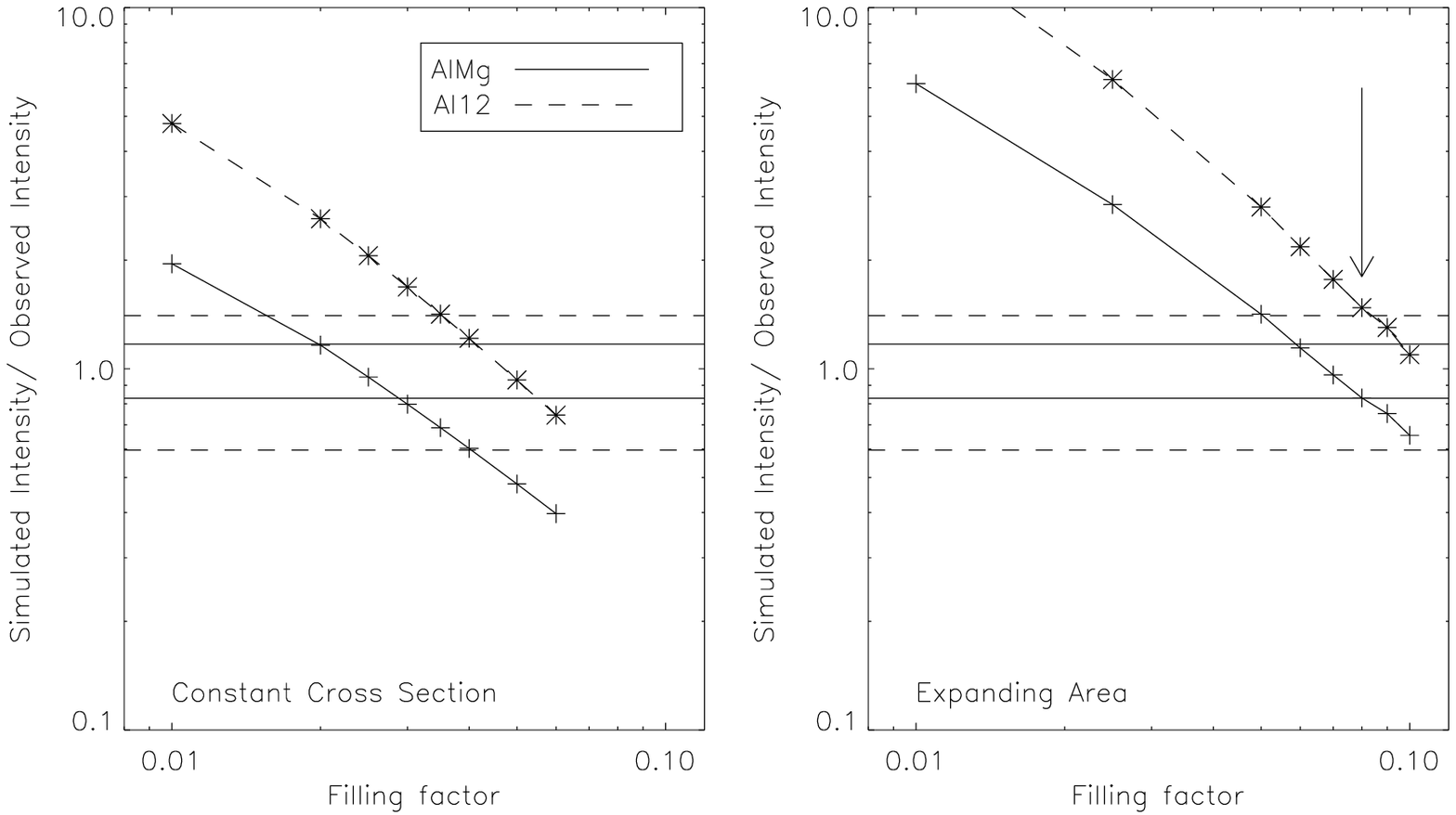}}}
\caption{The ratio between the simulated X-ray intensity and observed X-ray intensity is shown above for the SXT AlMg filter
(solid line with crosses) and the SXT Al12 filter (dashed line with asterisks).  The plot on the left shows the results
for the constant cross-section case, the plot on the right shows the results for the expanding area case.  The horizontal
lines represent two standard deviations in the AlMg intensity (solid) and Al12 intensity (dashed).  The best fit to the
X-ray intensities is a filling factor of 8\% in the expanding area case.
\label{fig:int_rat}}
\end{figure*}

Figure~\ref{fig:int_rat} shows ratio of the simulated SXT filter intensities to the observed filter
intensities as a function of filling factor for the two different geometry assumptions.  
The plot on the left assumes the loops in the ensemble have  constant cross sections, while the plot on the right assumes the 
area of each loop expands proportional to $B(s)^{-1}$.  
The AlMg filter ratio is shown as the 
solid line with crosses and the Al12 ratio is shown as the dashed line with asterisks.  
When the curve equals 1, the simulated intensity matches the observed intensity at that filling factor.
Horizontal lines show twice the standard deviation implied by the AlMg observations (solid) and Al12 observations
(dashed).   For a positive result to be found, the Al12 curve must be within the dashed horizontal lines at the
same filling factor as the AlMg curve is within the solid horizontal lines.
For the constant cross-section case, no filling factor returns a solution that is within two standard deviations
of the observed intensities at the same filling factor.  For the expanding area geometry, a filling factor of 8\% estimated 
the AlMg intensity at 2 standard deviations less than the observed intensity (80\% of the observed value) and a Al12 filter 
intensity at 2.3 standard deviations above the observed intensity (147\% of the observed value).  This is the best match
between the simulations and observations.

\begin{figure}[t!]
\centerline{
\includegraphics{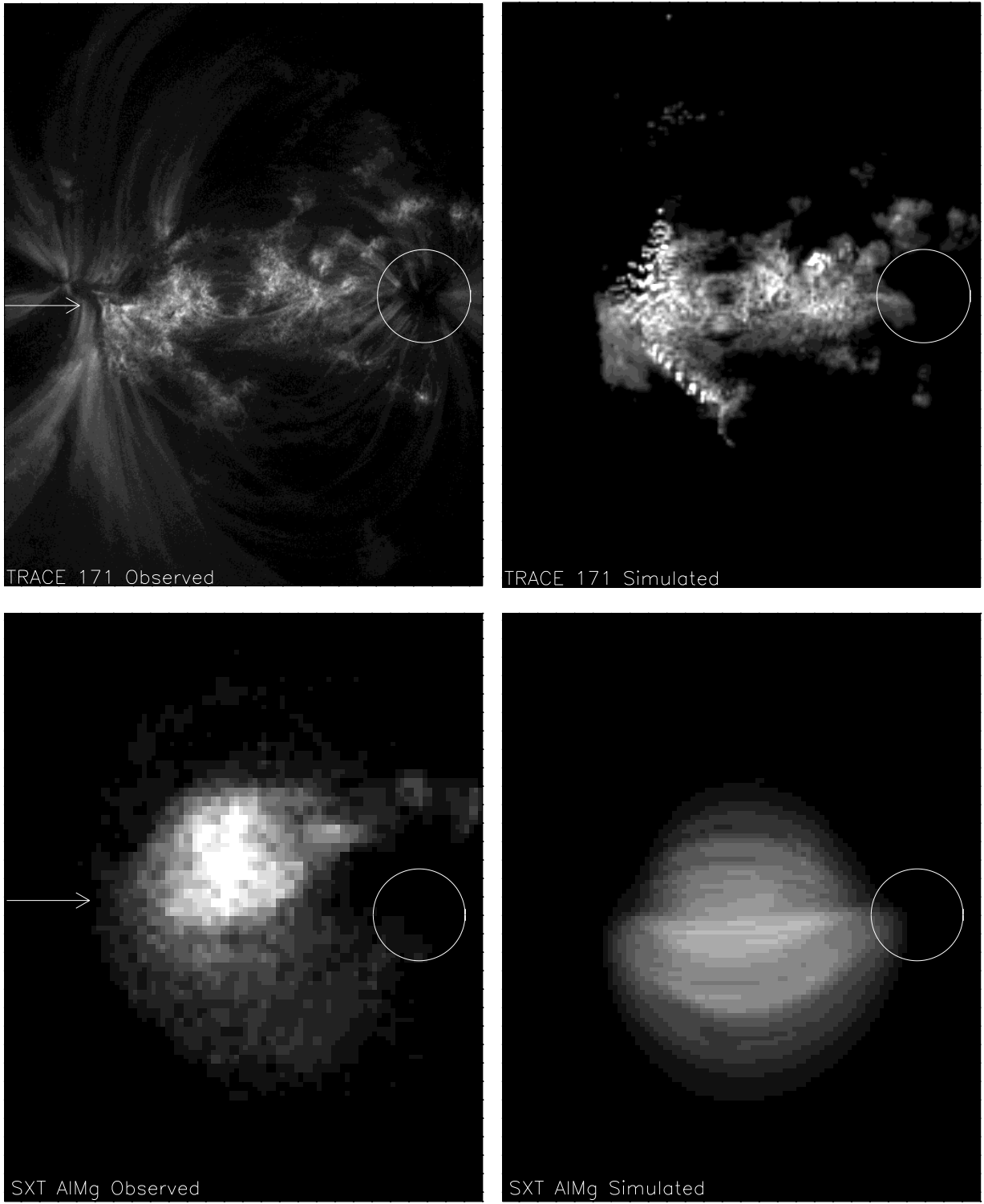}}
\caption{Left: The observed TRACE 171 and SXT AlMg images.  Right: The simulated TRACE 171 and SXT AlMg images for the
best fit case.  The images are displayed linearly and scaled identically.
\label{fig:comparison}}
\end{figure}

\begin{figure*}[t!]
\centerline{
\resizebox{18cm}{!}{\includegraphics{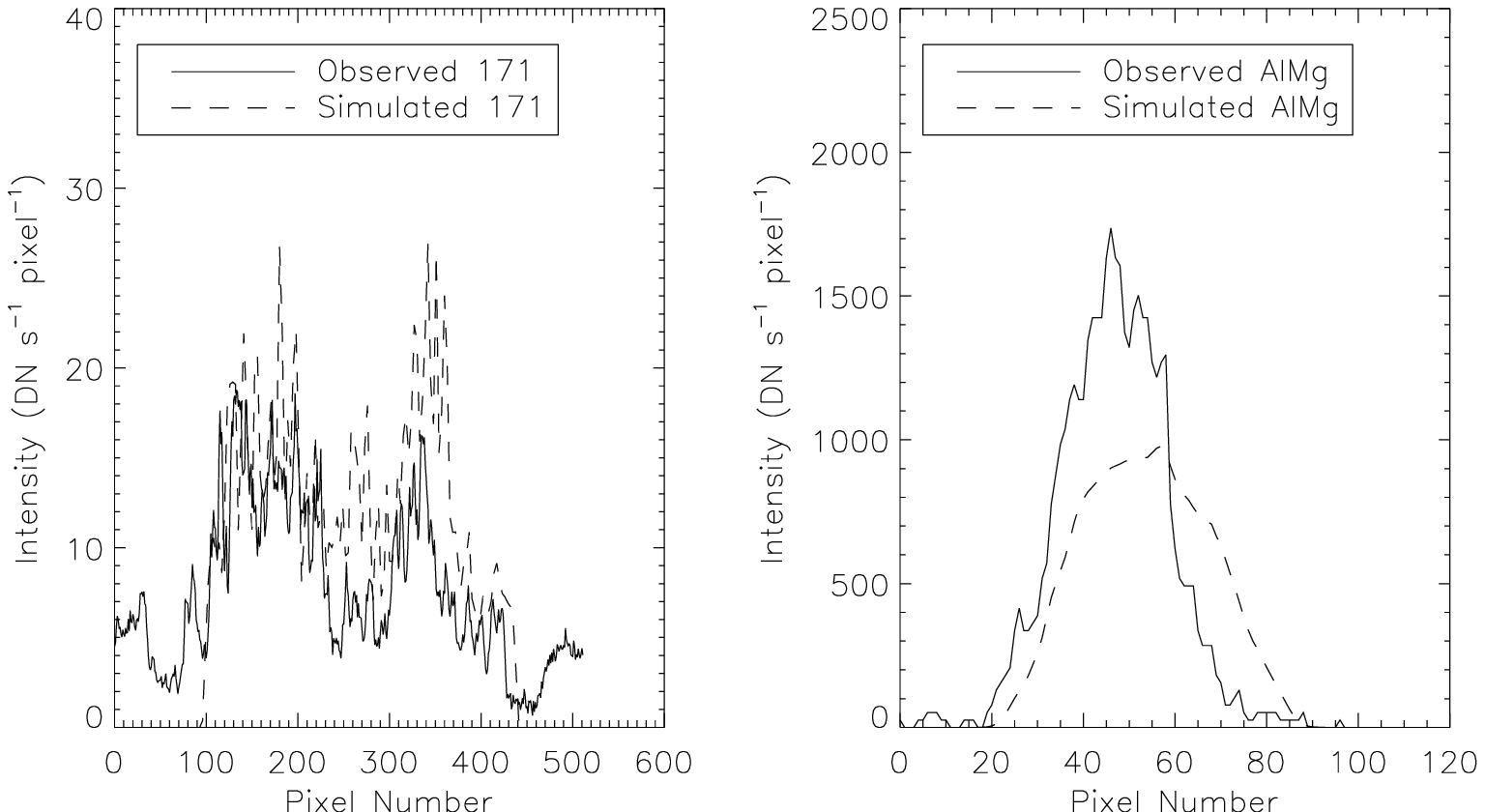}}}
\caption{A comparison of the simulated and observed TRACE 171 intensities (left) and SXT AlMg intensities (right) 
along a single horizontal cut.
\label{fig:lc}}
\end{figure*}

Figure~\ref{fig:comparison} shows the observed {\it TRACE} 171 and SXT AlMg images as well as the simulated images 
for the case that provides the best fit.  
The images are displayed linearly with identical scaling.  In all images, a circle is drawn that highlights a region
where the simulated and observed morphology differ.  In the observations in this region, there are bright extended EUV loop legs, 
but no moss.  There is also little SXT emission in this region. All the simulated images have both moss and SXT emission in the
region.   Additionally, there are several bright dots in the simulated EUV image that are not present in the 
observation.  These represent footpoint emission from a field line that originates in a moss region, but does
not terminate in a moss region region.
In Figure~\ref{fig:lc}, we show the intensity along a horizontal cut across both the TRACE 171 images and the SXT AlMg images.  
The two arrows in Figure~\ref{fig:comparison} indicate the vertical position of the cut.  
The SXT intensities are comparable
shapes, though the simulated intensity is less than the observed intensity and shifted to the right.


\section{Discussion}

\begin{figure*}[t!]
\centerline{
\resizebox{18cm}{!}{\includegraphics{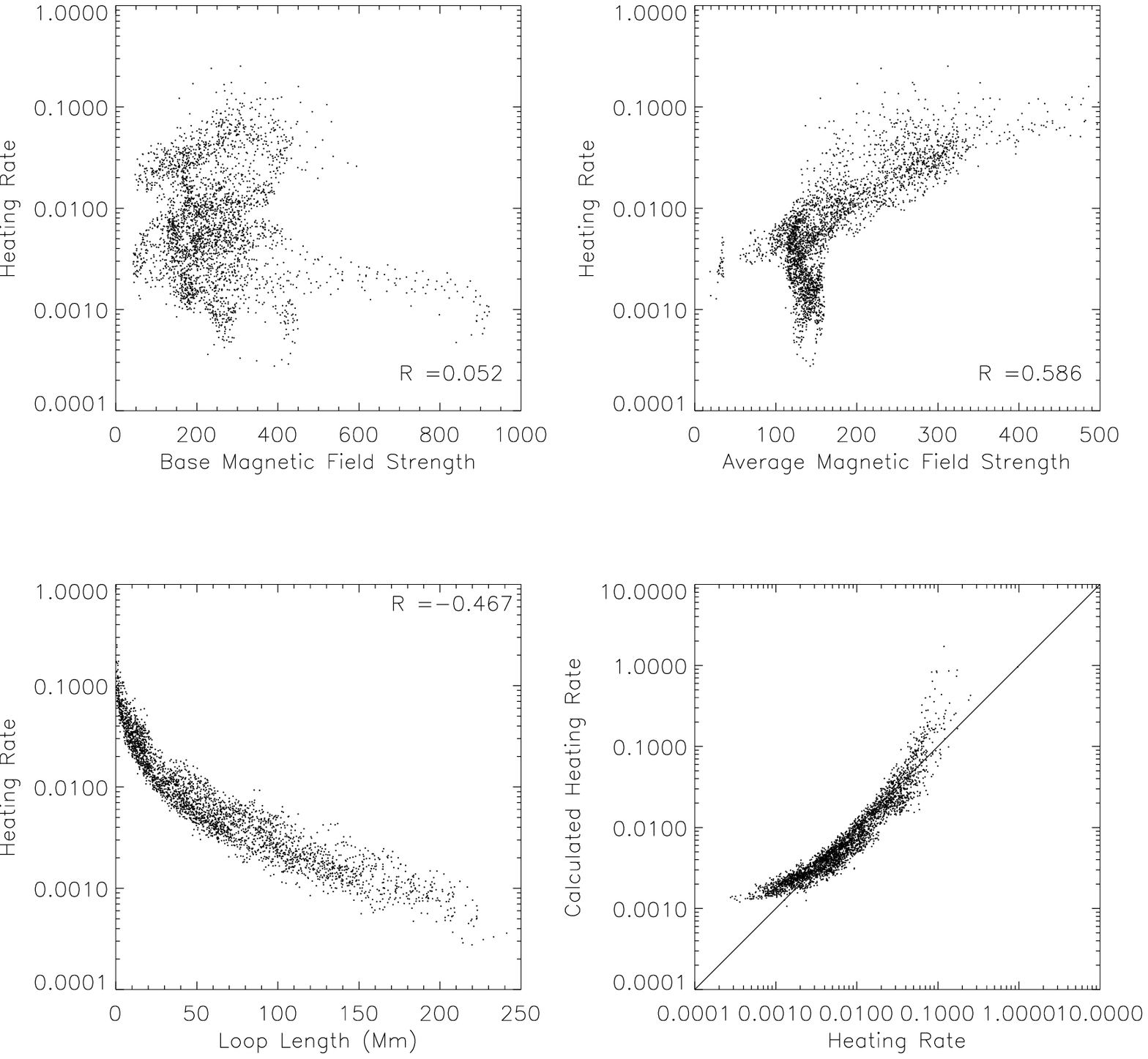}}}
\caption{The best heating rate for each field line as a function
of the base magnetic field strength (top left), average
magnetic field strength (top right), and loop length (bottom
left).  The heating rate calculated from the fit parameters
is also shown as a function of the best heating rate (bottom right).
\label{fig:ebl}}
\end{figure*}

In this paper, we have used a new technique to infer the steady heating rate from
the EUV moss intensities to test whether the loops in active region cores agree with the solutions
of the hydrodynamic equations for steady, uniform heating. 
We have simulated the core of Active Region 9017 using steady uniform heating
along potential field lines to match within 5\% the TRACE 171\,\AA\ moss intensities at
the footpoint.  We find the best match between the simulated and observed SXT AlMg and Al12 intensities
occurs at a filling factor of 8\% for loops that expand with height proportional to $B(s)^{-1}$.

Our intention with this research was to test whether the steady EUV moss and X-ray core
emission could self-consistently agree with ensembles of steadily heated loops.  
These results demonstrate that the intensities in these regions can, at best, be matched to steadily uniformly
heated loops within approximately two standard deviations in both SXT filters.  This disparity is acceptable
considering the systematic errors associated with this study, such as assuming the geometry of the 
field was well represented by the potential field and that all loops in the ensemble had the same
filling factor or cross sectional area.  Better agreement may have been possible if we had allowed
the heating to be non-uniform, but we did not examine this possibility.

Density measurements at the footpoints of the active region core loops would allow for the filling factor to be 
measured at each of the loop footpoints.  Knowing the filling factor, moss intensity, and loop length would fully constrain the 
steady uniform heating equations and provide a true test of the steady heating model.  This density measurement is now possible
with the EIS instrument on Hinode (\citealt{warren2007}).

Unlike previous studies of ensembles of loops which heated the loops based on an assumed heating rate proportional to 
$\bar{B}^\alpha/L^\beta$, we have calculated the heating rate along each individual field line 
based solely on the observed moss intensity and assumed filling factor.
We can now characterize the resulting relationships between the
calculated heating rate and other characteristics of the region, such as loop length and field strength.
Figure~\ref{fig:ebl} shows the correlation between the calculated heating rate and base field strength,
average field strength along the loop, and loop length.  Using a regression technique, we find that
the volumetric heating rate is best described by
$E = 0.051 (\bar{B}/B_0)^{0.29 \pm 0.03} (L/L_0)^{-0.95 \pm 0.01}$.  This calculated heating rate is shown as a function of the
real heating rate in the bottom right of Figure~\ref{fig:ebl}.  
In the above equations, $\bar{B}$ is the average magnetic field strength along the loop in Gauss, and $L$ is
the loop length in Mm and $B_0$ and $L_0$ are chosen to be 76 Gauss and 29 Mm respectively
to be comparable to previous work (\citealt{warren2006}). 
The correlation coefficient is
0.71 between the log of average field strength and the log of the heating rate and -.94 between the log of the
loop length and log of the heating rate.  There is no correlation between the heating rate and footpoint field 
strength (correlation coefficient = -0.047).  

The inverse relationship between the heating rate and loop length matches the results of the other forward
modeling studies (\citealt{schrijver2004,warren2006}).  However, the previous studies determined the best scaling
with the average magnetic field strength was for $1.0$, where we find a scaling of $0.3$.  This discrepancy is most
likely due to our strict matching of the TRACE moss intensity, where previous studies focused on matching the SXT
intensity and only did a rough comparison with the observed and simulated moss intensity.

In this study, we used potential field extrapolations to approximate the loop geometry and the
moss regions themselves to select the footpoints of the heated field lines.  However,
the resulting morphological
comparison of the observed images with the simulated images show some discrepancy.
Specifically, several field lines terminate in the negative polarity sunspot where there is no SXT or TRACE moss
emission observed;  this is shown with the circle in Figures~\ref{fig:multipanel} and \ref{fig:comparison}.  
Instead, there are several extended loop legs seen in the EUV.  This could be an indication that the 
connectivity of the field was not correct; however
the connectivity was not improved when using the vector field data or when considering linear force free
field extrapolations for different force free parameters.  
Another option is that those loops are heated asymmetrically causing moss on one side of the loop
and extended EUV emission on the other side of the loop.  Most coronal heating theories rely on photospheric
motions as a driving force. Because photospheric motions are suppressed in sunspot
regions, asymmetric heating along loops with one footpoint in a sunspot is a strong possibility.  
We did not consider asymmetric or non-uniform heating in this study.  In the future, it may be possible
to consider the EUV emission at both footpoints to fully limit the asymmetry of the heating function.

\cite{martens2000} derived the relationship between the base pressure, $p_0$, filling
factor, $f$, and TRACE 171 moss intensity, $I_{moss}$, i.e., $p_0 f = 0.050 I_{moss}$
using a simple radiative loss function and an analytical approximation to the hydrostatic
equations.  We derive this equation using the most recent radiative loss function (\citealt{brooks2006}).
 We find the relationship $p_0 f = 0.026 I_{moss}$
best represents the simulations.


\acknowledgments
ARW was supported by a NASA Sun-Earth Connection Guest Investigator grant
and NSF Career grant.



\end{document}